\documentclass[10pt, prl, aps, twocolumn, showpacs, citeautoscript, floatfix, reprint, amsmath, amssymb, notitlepage,superscriptaddress]{revtex4-1}

\usepackage{graphicx}
\usepackage{stmaryrd}
\usepackage{rotating}
\usepackage{wasysym}
\usepackage{amsmath}
\usepackage{amsfonts}
\usepackage{amssymb}
\usepackage[countmax]{subfloat}
\usepackage{dcolumn} % align table columns on decimal point
\usepackage{bm} % bold math
\usepackage{color}

\usepackage{float}
\usepackage{latexsym}
\usepackage[colorlinks, citecolor=blue, linkcolor=blue, urlcolor=blue]{hyperref}

\usepackage{color}

\begin{document}

\title{Renormalization group approach to symmetry protected topological phases}

\author{Evert P. L. van Nieuwenburg} 

\affiliation{Institut f\"{u}r Theoretische Physik, ETH-Z\"urich, 8093 Z\"urich, Switzerland}
\affiliation{Institute for Quantum Information and Matter, Caltech, Pasadena, 91125 CA, USA}

\author{Andreas P. Schnyder} 

\affiliation{Max-Planck-Institut f$\ddot{u}$r Festk$\ddot{o}$rperforschung, Heisenbergstrasse 1, D-70569 Stuttgart, Germany}

\author{Wei Chen} 

\affiliation{Institut f\"{u}r Theoretische Physik, ETH-Z\"urich, 8093 Z\"urich, Switzerland}
\affiliation{Department of Physics, PUC-Rio, Rio de Janeiro, Brazil}

\date{\rm\today}

\begin{abstract}

A defining feature of a symmetry protected topological phase (SPT) in one-dimension is the degeneracy of the Schmidt values for any given bipartition. For the system to go through a topological phase transition separating two SPTs, the Schmidt values must either split or cross at the critical point in order to change their degeneracies. A renormalization group (RG) approach based on this splitting or crossing is proposed, through which we obtain an RG flow that identifies the topological phase transitions in the parameter space. Our approach can be implemented numerically in an efficient manner, for example, using the matrix product state formalism, since only the largest first few Schmidt values need to be calculated with sufficient accuracy. Using several concrete models, we demonstrate that the critical points and fixed points of the RG flow coincide with the maxima and minima of the entanglement entropy, respectively, and the method can serve as a numerically efficient tool to analyze interacting SPTs in the parameter space.
 
\end{abstract}

%\pacs{64.60.ae, 64.60.F-}

%64.60.ae Renormalization-group theory

%64.60.F- Equilibrium properties near critical points, critical exponents

\maketitle

{\it Introduction.-} 
Symmetry protected topological phases (SPTs) in one-dimension (1D) have become 
one of the main focus of attention in the field of topologically ordered systems~\cite{Gu09,Senthil15}.
An SPT is a phase that  cannot be described by a local Landau order parameter, but instead is characterized by nonlocal topological properties
of the ground-state wave function, such as nonlocal string order~\cite{denNijs89,Kennedy92}, nonzero Berry phase~\cite{Hirano08}, or nontrivial quantum entanglement~\cite{ryu_hatsugai_PRB_06,matsuura_ryu_PRB_16}.  
These nonlocal properties can be conveniently calculated within the matrix product state (MPS) formalism~\cite{Fannes92} in combination with numerical methods, e.g., the infinite time-evolving block decimation (iTEBD) algorithm~\cite{Vidal07}.

Our focus in this paper is on the study of topological phase transitions between SPTs through the use of a renormalization group (RG) approach. This is motivated in part by recent progress in the development of an RG formalism for the phase transitions of topological insulators~\cite{Chen16,Chen16_2,Chen17}. 
Within this formalism, the RG scheme renormalizes either the Berry connection, the (many-body) Berry curvature, or the Pfaffian of the time reversal operator. In this way, a flow equation is derived which characterizes topological phase transitions in both noninteracting~\cite{Chen16,Chen16_2} and interacting~\cite{Chen17,Kourtis17,Chen18} systems.
This is particularly useful for interacting systems, where the RG scheme provides an efficient tool to determine topological phase transitions, since only very few points need to be computed numerically~\cite{Kourtis17,Chen18}. It is then intriguing to ask whether an RG scheme based on a similar principle can be constructed also for SPTs.

Here, we answer this question affirmatively, by presenting an RG scheme for SPTs which is based on the degeneracy of Schmidt values for a bipartition of the system~\cite{Pollmann10,Turner11,Pollmann12}. The degeneracy of the Schmidt values is a defining feature of SPTs, which reflects itself also in degeneracies of the entanglement spectrum~\cite{Li08,Levin06,Kitaev06,Fidkowski10,Turner10}. In this sense, the degeneracy pattern by itself is already sufficient for identifying different SPTs, and the RG scheme is a strategy to detect the boundaries in the parameter space at which the degeneracy pattern changes.

%The changes in these degeneracies provide a way of identifying phase transitions involving SPTs, and the RG scheme we develop takes advantage of this.

For concreteness, let us consider a Hamiltonian that describes different SPTs as a function of tuning parameters ${\bf M}=(M_{1},M_{2},...)$.
As ${\bf M}$ is tuned across a critical point ${\bf M}_c$
separating two phases (at least one of them being an SPT), the degeneracy pattern of the Schmidt values changes. Moreover, for the topological phase transition to be of second-order (meaning that the ground-state wave function continuously evolves through the critical point) the Schmidt values must either split or cross at the critical point ${\bf M}_c$ to do so. Based on this observation, we propose an RG scheme that allows to characterize phase transitions between SPTs in a numerically efficient manner, solely from the two largest Schmidt values, without explicitly invoking the symmetry that protects the SPTs. We demonstrate this method for two concrete models, namely the 
spin-1 Heisenberg antiferromagnetic chain~\cite{Haldane83,Haldane83_2} and the 
spin-1 two-leg spin ladder~\cite{takayama_spin_1_two_leg_ladder_PRB_01}. 
The proposed RG scheme is applicable to any one-dimensional SPT and can 
 be extended also to two or higher dimensions, e.g., by use of the iPEPS algorithm~\cite{cirac_iPEPS_PRL_08}.

\begin{figure}
\includegraphics[clip=true,width=0.99\columnwidth]{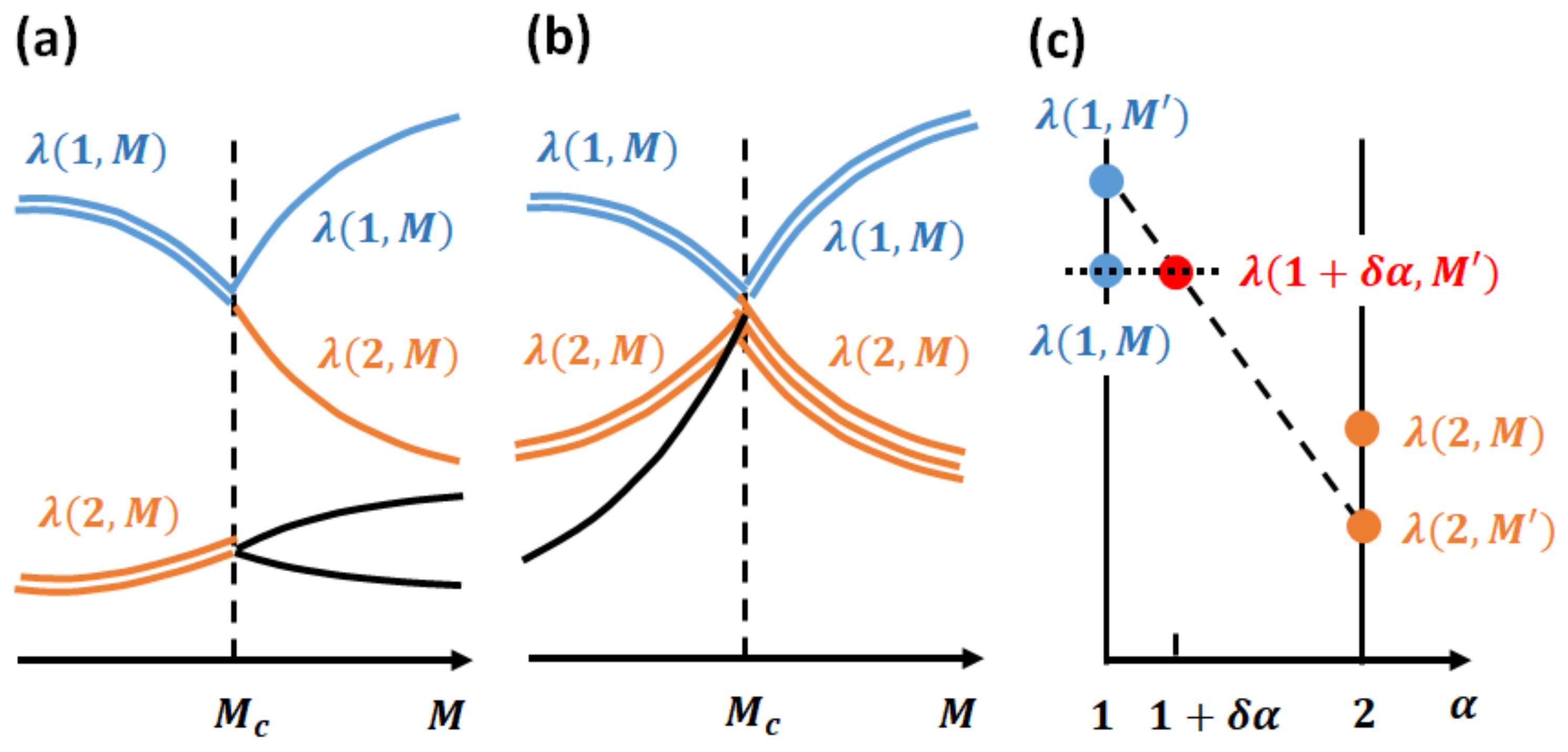}
\caption{(a) The ``split" scenario of degenerate Schmidt values, in which the doubly degenerate Schmidt values in the left phase (double lines)
split into non-degenerate Schmidt values in the right phase. (b) The ``cross" scenario, in which the Schmidt values in the left phase with degeneracies $(n_1, n_2 , n_3) = (2,2,1)$ and those in the right phase 
with degeneracies $(n_1, n_2) = (2,3)$ cross at the critical point ${\bf M}_c$. (c) Schematics of the SVRG approach. At a given ${\bf M}$, we seek for the new ${\bf M}^{\prime}$ that satisfies Eq.~(\ref{lambda_scaling_procedure}), as indicated by the dotted line. 
Here,  $\delta\alpha$ is a small positive number which serves as the scaling parameter, and $\lambda(1+\delta\alpha,{\bf M})$ is a linear interpolation function (dashed line).
  } 
\label{fig:SVRG_interpolate}
\end{figure}

%Our RG scheme is based on the MPS formalism that expresses the ground state of a 

{\it Schmidt decomposition and SPTs.-} 
Let us consider a gapped 1D system with tuning parameter ${\bf M}$, ground state  $|\psi\rangle$, and Hilbert space ${\cal H}$, which is partitioned into   ${\cal H} = {\cal H}_L \otimes {\cal H}_R$.
In the Schmidt decomposition  the ground state $|\psi\rangle$ is decomposed into  $|\psi\rangle=\sum_{\beta} \tilde{\Lambda}_{\beta} ({\bf M})  |\beta_{L}\rangle\otimes|\beta_{R}\rangle$,
where the Schmidt values $\tilde{\Lambda}_{\beta}  ({\bf M} )  \geq 0$ are positive and the Schmidt states $|\beta_{R,L}\rangle$ form orthogonal basis sets for ${\cal H}_{R,L}$~\cite{pollmann_les_houches}.
The Schmidt decomposition is directly related to the bipartite entanglement, since the Schmidt states are the eigenstates of the reduced density matrix 
$\rho_{{R}} ={\rm Tr}_{{L}} \left(|\psi\rangle\langle\psi|\right)$ and
the Schmidt values are the square root of the corresponding eigenvalues, i.e., 
 $\rho_{{R}} = \sum_{\beta }   \tilde{\Lambda}^2_{\beta} ({\bf M}) |\beta_{R}\rangle\langle\beta_{R}|$.
 Thus, the entanglement entropy $S({\bf M})$ and the entanglement spectrum $\epsilon(\beta,{\bf M})$ can be expressed in terms of $\tilde{\Lambda}^2_{\beta} ({\bf M})$ via
 $S({\bf M})=-\sum_{\beta} \tilde{\Lambda}^2_{\beta} ({\bf M})  \ln \tilde{\Lambda}^2_{\beta} ({\bf M})$ and   $\epsilon(\beta,{\bf M})=-\ln \tilde{\Lambda}^2_{\beta} ({\bf M})$, respectively. 
For the sake of normalization [see Eq.~(\ref{sum_schmidt})], we study in the following the renormalization of the \emph{square} of the Schmidt values, i.e., 
$\tilde{\lambda} ( \beta, {\bf M} ) = \tilde{\Lambda}^2_{\beta} ({\bf M})$, and assume
that these are enumerated in descending order, i.e., $\tilde{\lambda}(1,{\bf M})\geq\tilde{\lambda}(2,{\bf M})\geq\tilde{\lambda}(3,{\bf M})\geq...\geq\tilde{\lambda}(N_{R},{\bf M})$. 

The Schmidt decomposition and, in particular, the degeneracies of the Schmidt values contain crucial information about the topology of the ground state~\cite{pollmann_les_houches}. That is,
all Schmidt values of an SPT are necessarily even-multiplicity degenerate, which is guaranteed by the defining symmetries of the SPT. Moreover, 
different SPTs can be distinguished by the projective representations under which the Schmidt states transform~\cite{Turner11,fidkowski_kitaev_projective_reps_1D_PRB_11}. 
Therefore, at a critical point ${\bf M}_c$ in between two phases, the degeneracy pattern of the Schmidt values must change. 
In the following we use these observations to construct a topological invariant from the Schmidt values and to derive an RG scheme for phase transitions in SPTs.

{\it Renormalization group procedure.-} 
In general we label the degeneracies of a given set of Schmidt values in terms of a vector $\left\{n_{\alpha}\right\}=\left\{n_{1},n_{2},n_{3}...\right\}$, with each $n_{\alpha} \geq 1$ representing the multiplicity. In the case of an SPT, each of the $n_{\alpha}$ is a multiple of $2$.
A complete calculation of $\left\{n_{\alpha}\right\}$ can be achieved for any SPT, for example, by use of Young tableaux~\cite{Morimoto14}, but this shall not
be our concern. For the purpose of the RG scheme, we consider the set of $\lambda(\alpha,{\bf M})$ with the degeneracies removed~\cite{footnote1}. 
That is, the $\lambda(\alpha,{\bf M})$ are unique for different $\alpha$, are labelled in descending order
\begin{eqnarray}
\lambda(1,{\bf M})>\lambda(2,{\bf M})>\lambda(3,{\bf M})>...>\lambda(k,{\bf M})\; ,
\label{sort_lambda}
\end{eqnarray}
 and satisfy 
 \begin{eqnarray} \label{sum_schmidt}
&&\sum_{\alpha}n_{\alpha}\lambda(\alpha,{\bf M})=\sum_{\beta=1}^{\chi}\tilde{\lambda}(\beta,{\bf M})=1.
\end{eqnarray}
The integer $\chi$ in the above equation denotes the total number of Schmidt values, which, when computed within the MPS framework, corresponds to the required bond dimension of an exact MPS. The $\left\{n_{\alpha}\right\}$ in Eq.~\eqref{sum_schmidt} can be viewed as a topological invariant, since it does not change under smooth deformations within a given SPT. Two distinct SPTs are characterized by two different sets of $\left\{n_{\alpha}\right\}$.

The Schmidt values must either split or cross at the critical point ${\bf M}_c$ between two phases in order to reshuffle the degeneracy pattern $\left\{n_{\alpha}\right\}$, such that a topological phase transition can take place, as shown schematically in Figs.~\ref{fig:SVRG_interpolate}(a) and~\ref{fig:SVRG_interpolate}(b). We propose the following RG scheme to capture the splitting or crossing of Schmidt values without changing the topological invariant $\left\{n_{\alpha}\right\}$. Starting from a given value of the tuning parameter ${\bf M}$, we search for a new value ${\bf M}'$ that satisfies
\begin{eqnarray}
\lambda(1,{\bf M})=\lambda(1+\delta\alpha,{\bf M}^{\prime}) ,
\label{lambda_scaling_procedure}
\end{eqnarray} 
where $\delta\alpha >0$  is the scaling parameter. 
Here, $\lambda(1+\delta\alpha,{\bf M}^{\prime})$ is taken to be a function that linearly interpolates between $\lambda(1,{\bf M}^{\prime})$ and $\lambda(2,{\bf M}^{\prime})$.
That is, Eq.~\eqref{lambda_scaling_procedure}  demands that the first Schmidt value $\lambda(1, {\bf M} )$ at a given ${\bf M}$ is equal to the interpolated Schmidt value 
$\lambda(1+\delta\alpha, {\bf M}' )$ at the new ${\bf M}^{\prime}$, as indicated schematically in Fig.~\ref{fig:SVRG_interpolate}(c). 
Iteratively solving the mapping ${\bf M}\rightarrow{\bf M}^{\prime}$ yields an RG flow in the parameter space of ${\bf M}$. Expanding Eq.~\eqref{lambda_scaling_procedure} in both $d {\bf M} = {\bf M}' - {\bf M}$ and $dl = \delta \alpha$ yields the leading order RG equation  
\begin{subequations} \label{both_RG_equations}
\begin{eqnarray}
\frac{dM_{i}}{dl}=-\frac{\partial_{\alpha}\lambda(\alpha,{\bf M})|_{\alpha=1}}{\partial_{M_{i}}\lambda(1,{\bf M})}\;,
\label{RG_eq_generic}
\end{eqnarray}
for each component $M_i$ of the tuning parameter. For numerical simulations, it is useful to rewrite Eq.~\eqref{RG_eq_generic} in discretized form 
\begin{eqnarray}
\frac{dM_{i}}{dl}=\frac{\Delta M_{i}}{\Delta\alpha}\frac{\lambda(1,{\bf M})-\lambda(2,{\bf M})}{\lambda(1,{\bf M}+\Delta M_{i}{\hat{\bf M}_{i}})-\lambda(1,{\bf M})}\;,
\label{RG_eq_numerical}
\end{eqnarray}  
\end{subequations}
where $\Delta M_{i}$ is the numerical grid spacing along the ${\hat{\bf M}}_{i}$ direction and $\Delta\alpha=2-1=1$. For the concrete models described below, we implement Eq.~(\ref{RG_eq_numerical}) using the MPS formalism together with the iTEBD algorithm.

Equations~\eqref{both_RG_equations} are the main results of this article. We call this RG scheme ``Schmidt value renormalization group" (SVRG), since the RG flow is obtained from renormalizing the Schmidt values.
This approach allows to efficiently characterize phase transitions between SPTs, since only the largest two degeneracy-removed Schmidt values $\lambda(1,{\bf M})$ and $\lambda(2,{\bf M})$
are needed. With the knowledge of the first two degeneracy patterns $\left\{n_{1},n_{2}\right\}$ at hand, the approach is a rigorous method to detect the phase boundary at which $\left\{n_{1},n_{2}\right\}$ changes in the ${\bf M}$ parameter space. The approach becomes particularly useful when it is implemented within the MPS framework, where $\lambda(1,{\bf M})$ and $\lambda(2,{\bf M})$ can be computed to high accuracy already
for relatively small bond dimensions $\chi$. The obtained phase boundary is only as accurate as that with which the largest Schmidt values are obtained. It should be noted that in practice it may not be known \emph{a priori} what the degeneracy pattern $\left\{n_{\alpha}\right\}$ of the Schmidt values are. Hence to capture at least the largest two sets of degeneracies $\left\{\lambda(1,{\bf M}),\lambda(2,{\bf M})\right\}$ it is not sufficient in general to use a bond dimension $\chi$ equal to their total multiplicity. The numerical effort required is therefore similar to performing a full iTEBD simulation. However, due to the flow diagram from the RG procedure, one is no longer required to sample a fine grid in parameter space to locate the phase boundary.

{\it Critical points and fixed points.-} The principle behind Eq.~(\ref{lambda_scaling_procedure}) is that it reduces the deviation of Schmidt values away from their fixed point configuration. This can be seen by analyzing the RG flow in the parameter space of ${\bf M}$ described by Eqs.~\eqref{both_RG_equations}. The fixed points of this RG flow
correspond to points/lines in the parameter space, where the separation between the largest two Schmidt values $\lambda ( 1, {\bf M})$
and $\lambda (2, {\bf M})$ are extremal. To make this more precise, 
let us consider $\lambda ( \alpha, {\bf M})$ close to a fixed point ${\bf M}_f$ and study its deviation away from ${\bf M}_f$,
i.e.,  
$\lambda_{v}(\alpha,{\bf M}) = \lambda(\alpha,{\bf M}) - \lambda_{f}(\alpha,{\bf M}_{f})$, 
where $\lambda_{f}(\alpha,{\bf M}_{f})$ is the Schmidt value at the fixed point. 
Since $\lambda_{f}(\alpha,{\bf M}_{f})$ is invariant under the operation of Eq.~(\ref{lambda_scaling_procedure}), the deviation part itself satisfies $\lambda_{v}(1,{\bf M})=\lambda_{v}(1+\delta\alpha,{\bf M}^{\prime})=\lambda_{v}(1,{\bf M}^{\prime})+\delta\alpha\partial_{\alpha}\lambda(\alpha,{\bf M}^{\prime})|_{\alpha=1}$. Using the linear interpolation function, cf.~Fig.~\ref{fig:SVRG_interpolate}(c), we obtain
\begin{eqnarray}
\delta\alpha=\frac{\lambda(1,{\bf M}^{\prime})-\lambda(1,{\bf M})}{\lambda(1,{\bf M}^{\prime})-\lambda(2,{\bf M}^{\prime})}\;,
\label{delta_alpha_resuit}
\end{eqnarray}
whose denominator is always positive because of Eq.~(\ref{sort_lambda}). Since $\delta\alpha$ is positive by definition, the numerator of Eq.~(\ref{delta_alpha_resuit}) is positive, $\lambda_{v}(1,{\bf M}^{\prime})-\lambda_{v}(1,{\bf M})>0$. In other words, along the RG flow ${\bf M}\rightarrow{\bf M}^{\prime}$, $\lambda_{v}(1,{\bf M}^{\prime})-\lambda_{v}(1,{\bf M})>0$ is always satisfied. 
On the other hand, as we approach the critical point ${\bf M}\rightarrow{\bf M}_{c}$, the largest Schmidt value $\lambda(1,{\bf M})$ must decrease in order to meet with the second Schmidt value $\lambda(2,{\bf M})$, indicating $\lambda_{v}(1,{\bf M})<0$. Combining the above two inequalities yields  $|\lambda_{v}(1,{\bf M}^{\prime})|<|\lambda_{v}(1,{\bf M})|$, i.e., the magnitude of the deviation away from the fixed point configuration is gradually reduced under this scaling procedure, and hence the system gradually flows away from the phase boundary towards the fixed point.

%The exception to this occurs when a degenerate $\lambda(1,{\bf M})$ splits instead of meeting with $\lambda(2,{\bf M})$, a scenario we will call the split scenario as opposed to the cross scenario [c.f. Figs.~\ref{fig:SVRG_interpolate}(a) and~\ref{fig:SVRG_interpolate}(b)].

{\it Split and cross scenarios.-} There are two scenarios for how the first two degeneracy patterns $\left\{n_{1},n_{2}\right\}$ can change at a phase boundary ${\bf M}_c$. For the numerical simulations it is important to distinguish between them. 

%There are two different possibilities for how the largest two $\lambda(\alpha,{\bf M})$ can change their degeneracies at ${\bf M}_c$, i.e., at an unstable fixed point.

\paragraph{1.~Split scenario.} In the split scenario, depicted in Fig.~\ref{fig:SVRG_interpolate}(a), $\lambda(1,{\bf M})-\lambda(2,{\bf M})$ is finite as the system approaches the critical point from one side ${\bf M}\rightarrow{\bf M}_{c}^{+}$, while it is zero, as ${\bf M}_{c}$ is approached from the other side ${\bf M}\rightarrow{\bf M}_{c}^{-}$. Hence, according to Eqs.~\eqref{both_RG_equations},
%(\ref{RG_eq_generic}) and (\ref{RG_eq_numerical}), 
the flow rate of the RG flow satisfies
\begin{subequations} \label{both_conditions_for_critical_point}
\begin{eqnarray}
{\rm \emph{Split}:}\;\;&&\lim_{{\bf M}\rightarrow{\bf M}_{c}^{+}}\left|\frac{d{\bf M}}{dl}\right|={\rm finite}\;,\;\;\;\lim_{{\bf M}\rightarrow{\bf M}_{c}^{-}}\left|\frac{d{\bf M}}{dl}\right|=0\;.\;\;\;\;
\label{split_critical_point}
\end{eqnarray}
\paragraph{2.~Cross scenario.}
In the cross scenario, depicted in Fig.~\ref{fig:SVRG_interpolate}(b), $\lambda(1,{\bf M})-\lambda(2,{\bf M})$ is zero for both ${\bf M}\rightarrow{\bf M}_{c}^{+}$ and ${\bf M}\rightarrow{\bf M}_{c}^{-}$. Hence, the flow rate satisfies 
\begin{eqnarray}
{\rm \emph{Cross}:}\;\;\lim_{{\bf M}\rightarrow\left\{{\bf M}_{c}^{+},{\bf M}_{c}^{-}\right\}}\left|\frac{d{\bf M}}{dl}\right|=0\;.
\qquad \qquad \qquad \quad \quad 
\label{merge_critical_point}
\end{eqnarray}
\end{subequations}

At a stable fixed point ${\bf M}_{f}$, on the other hand, the flow rate obeys 
\begin{eqnarray}
\lim_{{\bf M}\rightarrow{\bf M}_{f}^{+}}\frac{d{\bf M}}{dl}=-\lim_{{\bf M}\rightarrow{\bf M}_{f}^{-}}\frac{d{\bf M}}{dl}=\pm\infty\; ,
\label{fixed_point}
\end{eqnarray}
which follows from $\lim_{{\bf M}\rightarrow{\bf M}_{f}}\partial_{M_{i}}\lambda(1,{\bf M})=0$ together with Eq.~\eqref{RG_eq_generic}.
%The  fixed point is characterized by zero slope of the largest degeneracy-removed Schmidt value $\lim_{{\bf M}\rightarrow{\bf M}_{f}}\partial_{M_{i}}\lambda(1,{\bf M})=0$, therefore  
In numerical simulations, the above three equations, together with the knowledge of whether $\left\{n_{1},n_{2}\right\}$ has been changed, can be used to identify the critical points and fixed points in the phase diagram of a given  1D system.

Before discussing two applications of our SVRG approach, some remarks are in order. 

\textit{(i)} 
Eqs.~\eqref{both_conditions_for_critical_point} are necessary but not sufficient conditions for a topological phase transition. 
This is because Schmidt values can split and cross also at other types of phase transitions, for example, at a second-order quantum phase transition separating a magnetically ordered
state from a quantum disordered state.
Moreover, the Schmidt values can cross accidentally at ${\bf M}_{c}$, without an accompanying phase transition, in which case the degeneracy pattern simply swaps $\left\{n_1, n_2\right\} \to \left\{n_2, n_1\right\}$.
In order to confirm whether ${\bf M}_{c}$ is a phase transition point one needs to keep track of the entanglement entropy, since a maximum in this quantity signals a bulk transition. To identify whether a transition is topological, looking at the even-oddness of $\left\{n_{\alpha}\right\}$ is sufficient. It is of course also possible to check this by e.g. computing the Berry phase~\cite{Pollmann10}.

%An unstable fixed point ${\bf M}_{c}$ corresponds to a phase transition point only if it is accompanied by a maximum in the entanglement entropy, which signals a bulk phase transition. Hence, in order to confirm whether ${\bf M}_{c}$ is a phase transition point and whether it separates two topological phases, one needs to calculate also the entanglement entropy and the topology of the phases, e.g., by computing the Berry phase~\cite{Pollmann10}.

%Despite this caveat, it is generally not a difficult task because all ${\bf M}$'s inside a phase have the same topology, so one only needs to randomly choose one ${\bf M}$ to confirm the topology of the entire phase.

\begin{figure}
\begin{center}
\includegraphics[clip=true,width=0.99\columnwidth]{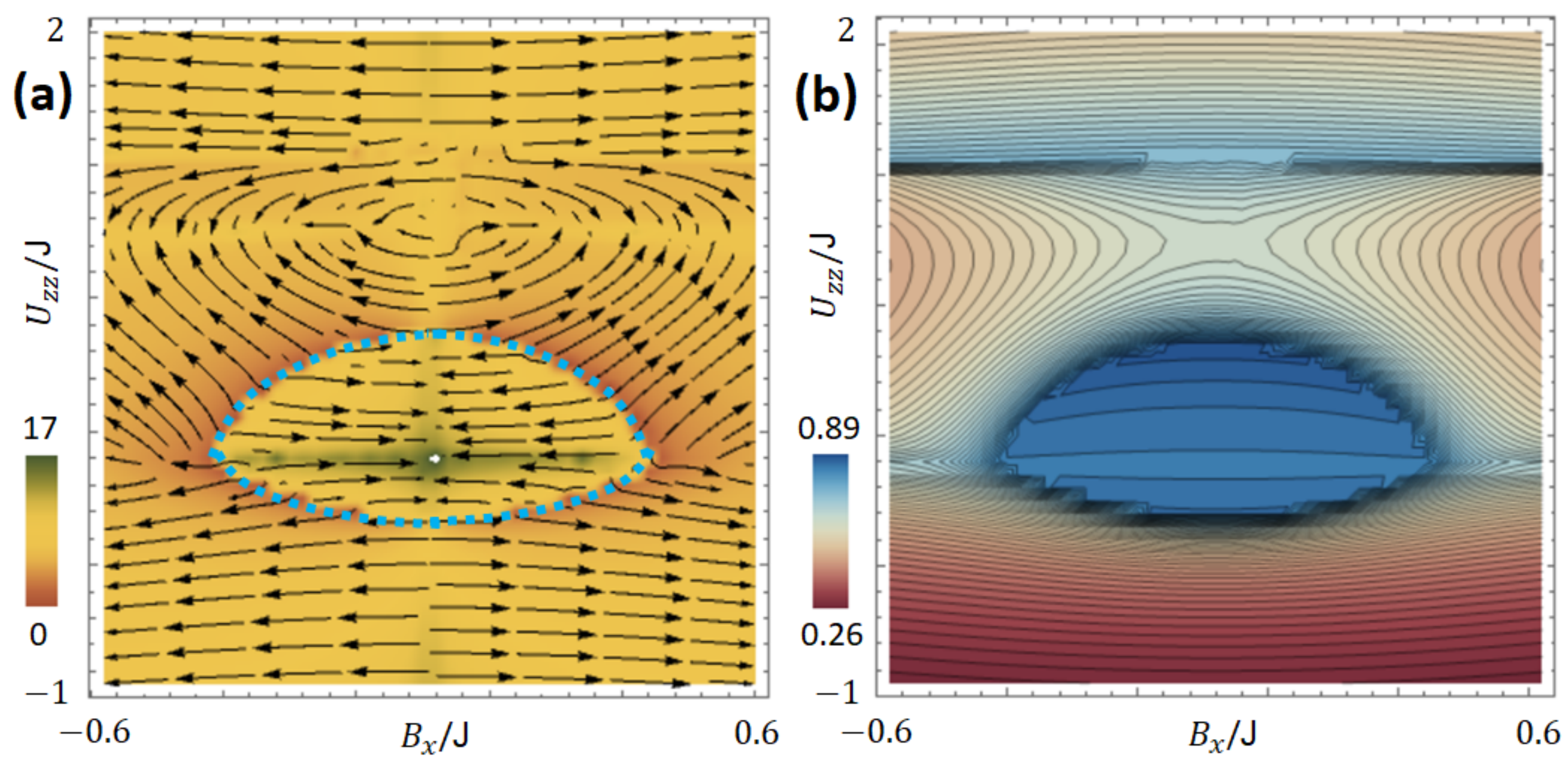}
\caption{(a) The RG flow (arrows) of the 
spin-1 Heisenberg antiferromagnetic chain, Eq.~(\ref{Haldane_chain}), in the parameter space ${\bf M}=(B_{x},U_{zz})$, obtained by means of the MPS simulations with bond dimension $\chi = 8$. 
The color scale indicates the flow rate in log scale. 
The RG flow reveals a topological phase transition indicated by the blue dotted line, which separates the topological
 Haldane phase (inside region) from topologically trivial phases (outside region). The SVRG correctly captures the SPTs of the model, but yields imprecise phase boundaries due to the small bond dimension.
% is restricted to incorrect phase boundaries by the small
%bond dimension.
 (b) Color-scale plot of the entanglement entropy of the spin-1 Heisenberg antiferromagnetic chain. }
\label{fig:Haldane_model_BxUzz}
\end{center}
\end{figure}

\begin{figure}
\begin{center}
\includegraphics[clip=true,width=0.99\columnwidth]{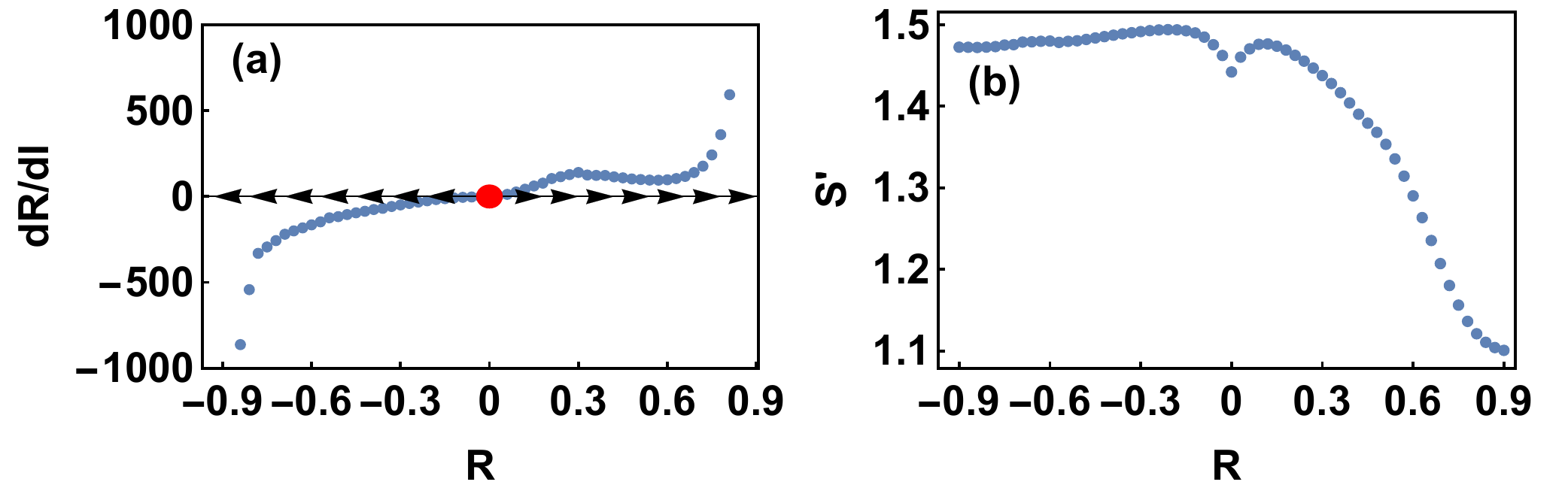}
\caption{Numerical results for the spin-1 two-leg spin ladder, Eq.~\eqref{ham_spin_ladder}, 
computed with MPSs of bond dimension $\chi=80$. (a) Interpolated Schmidt value RG flow in units of $\Delta R/\Delta\alpha=0.03$. (b) Entanglement entropy  $S$ contributed by the 8 largest Schmidt values. } 
\label{fig:spin1_ladder_cutleg}
\end{center}
\end{figure}

\textit{(ii)} Typically, the critical points and fixed points of the SVRG flow coincide with maxima or minima of the entanglement entropy $S$ in the parameter space {\bf M}, respectively. Below, we demonstrate this numerically for two examples, see Figs.~\ref{fig:Haldane_model_BxUzz} and~\ref{fig:spin1_ladder_cutleg}. 
For the case of critical points, this connection is obvious, since at the phase transition point the bulk gap closes, leading
to a diverging $S$. On the other hand, we observe from our numerical simulation that stable fixed points always correspond to the minimum in the entanglement entropy $S$. We can obtain further insight into this from the formula
\begin{eqnarray}
\partial_{M_{i}}S=-\sum_{\beta}\left[1+\ln\tilde{\lambda}(\beta,{\bf M})\right]\partial_{M_{i}}\tilde{\lambda}(\beta,{\bf M})\;.
\label{dSdM_minimum}
\end{eqnarray}
For the examples presented below, we empirically observe that the minimum of entanglement entropy $\partial_{M_{i}}S=0$ occurs when, in accordance with Eq.~(\ref{dSdM_minimum}), either (i) $\partial_{M_{i}}\lambda(\alpha,{\bf M})=0$ for some $\alpha$, which agrees with Eqs.~(\ref{RG_eq_generic}) and (\ref{fixed_point}), or (ii) two Schmidt values satisfy $\lambda(\alpha_{1},{\bf M})=\lambda(\alpha_{2},{\bf M})$ but $\partial_{M_{i}}\lambda(\alpha_{1},{\bf M})=-\partial_{M_{i}}\lambda(\alpha_{2},{\bf M})$, such that the two contributions to Eq.~(\ref{dSdM_minimum}) cancel.

{\it Applications.-} 
An example for the split scenario is the spin-1 Heisenberg antiferromagnetic chain~\cite{Haldane83,Haldane83_2} in the presence of a magnetic field~\cite{Pollmann10}, described by
\begin{eqnarray}
H=J\sum_{i}{\bf S}_{i}\cdot{\bf S}_{i+1}+B_{x}\sum_{i}S_{i}^{x}+U_{zz}\sum_{i}\left(S_{i}^{z}\right)^{2},
\label{Haldane_chain}
\end{eqnarray}
where $J>0$ is the exchange coupling, $B_{x}$ is the magnetic field, and $U_{zz}$ is an on-site spin anisotropy. 
To compute the Schmidt values and their RG flow we use the MPS framework with bond dimension $\chi=8$~\cite{Vidal07}. 
In Fig.~\ref{fig:Haldane_model_BxUzz} we show the Schmidt value flow in the parameter space ${\bf M}=(B_{x},U_{zz})$, which reveals a topological phase transition (blue dots), i.e., an unstable fixed line form which the arrows point away. 
This phase boundary separates the topological Haldane phase (inside) from
topologically trivial phases (outside). 
The phase transition is of the ``split" type, since 
the flow rate is zero (brown color scale) as the critical line is approached from the outside, while
it is finite (yellow color scale) as the phase boundary is approached from the inside [see Fig.~\ref{fig:SVRG_interpolate}(a) and Eq.~(\ref{split_critical_point})].
%At the phase transition line the entanglement entropy diverges, see Fig.~\ref{Haldane_chain}(b).  
The phase transition line in Fig.~\ref{fig:Haldane_model_BxUzz}(a) corresponds to a line of maxima in the entanglement entropy shown in Fig.~\ref{fig:Haldane_model_BxUzz}(b).  
Note that at the center of the topological Haldane phase there is an attractive fixed point, which coincides with a minimum in the entanglement entropy.

We observe that the SVRG approach correctly captures the distinct SPTs of this model~\cite{Pollmann10,Gu09}, even for a small number of Schmidt values (i.e., a small MPS bond dimension $\chi$), which clearly demonstrates the high numerical efficiency of this approach. Having said that, it should be noted that for such a small $\chi$, the precise \emph{location} of the phase boundaries is not captured accurately, which nevertheless improves as increasing $\chi$.

An example of the cross scenario without topological phase transitions is the spin-1 two-leg spin ladder described by~\cite{takayama_spin_1_two_leg_ladder_PRB_01,Pollmann12}
\begin{eqnarray} \label{ham_spin_ladder}
H&=&J_{\rm leg}\sum_{\gamma, i}{\bf S}_{\gamma,i}\cdot{\bf S}_{\gamma,i+1}
+J_{\rm rung}\sum_{i}{\bf S}_{1,i}\cdot{\bf S}_{2,i}\; ,
\end{eqnarray}
where ${\bf S}_{\gamma,i}$ denotes the spin operator on leg $\gamma=1,2$ at position $i$ 
and $J_{\rm leg}$ ($J_{\rm rung}$) denotes the
exchange coupling along legs (rungs). 
For the computation of the Schmidt values we use the MPS formalism, as before, with bond dimension $\chi = 80$.
The RG flow of the Schmidt values is presented in Fig.~\ref{fig:spin1_ladder_cutleg}(a) as a function of 
$R=J_{\rm rung}/(J_{\rm leg}+|J_{\rm rung}|)$, with $J_{\rm leg}>0$.
We observe that there exists an unstable fixed point at $R_{c}=0$, where the flow rate vanishes. 
This fixed point does not correspond to a (topological) phase transition, since it is not
accompanied by a maximum in the entanglement entropy [see Fig.~\ref{fig:spin1_ladder_cutleg}(b)] and does not involve even-multiplicity degeneracies.
Instead, the fixed point at $R_c$ represents an accidental crossing of Schmidt values without a phase transition (i.e., the system remains gapped)~\cite{Pollmann12,takayama_spin_1_two_leg_ladder_PRB_01}, in accordance with remark \textit{(i)}. As expected for the cross scenario [cf.~Eq.~\eqref{merge_critical_point}],
the flow rate $d R / d l$ vanishes on both sides of $R_c$.
 %As seen in Figs.~\ref{fig:spin1_ladder_cutleg}(a) and~\ref{fig:spin1_ladder_cutleg}(b),
In closing, we observe that at the attractive fixed points
  $R = \pm 1$ the entanglement entropy has a mimimum.
This reinforces our conjecture that attractive SVRG fixed points are  always accompanied by minima in the entanglement spectrum.

% and the divergence of RG flow as $R\rightarrow\pm 1$ again demonstrates that the fixed point coincides with the minimum of entanglement entropy in the $R$ parameter space.
%We use the largest $8$ Schmidt values to calculate the entanglement entropy, which decreases as $R$ approaches the product states at $R=1$ and $R=-1$. 

% It has been argued, though, that despite the merge of Schmidt values at $R_{c}=0$, it is not a topological phase transition\cite{Pollmann12}. Thus, to confirm whether ${\bf M}_{c}$ is a topological phase transition, it is important to check the topology of the two phases separated by ${\bf M}_{c}$, which is nevertheless a simple task in this case since the fixed point is a simple product state.

{\it Conclusions.-} 
In summary, we have proposed an RG scheme to study phase transitions between SPTs based on the degeneracy of Schmidt values for a bipartition of the system. Through renormalizing the Schmidt values, we have derived an RG flow that identifies the topological phase transitions in the parameter space. This
Schmidt value RG approach can be implemented numerically
in an efficient manner within the MPS formalism, as demonstrated for two concrete models. Due to its numerical efficiency, we
anticipate that this method can serve as a powerful tool
to search for, and analyze, distinct SPT phases in a large
parameter space of any given model.

%With these unique features and numerical advantages, we anticipate that SVRG can be a numerically powerful tool to study topological phase transitions between SPTs.
%
%As the method requires only the few largest Schmidt values, the method implemented by MPS with a small bond dimension $\chi$ renders a numerically efficient tool to search for distinct SPTs in a large parameter space. 

%
%the SVRG approach is proposed to investigate the topological phase transitions between SPTs. The approach is purely based on the degeneracy of Schmidt values, independent from the symmetry that protects the SPT. 
%
%Rigorously, SVRG is a method to capture the split or merge of Schmidt values at a second-order quantum phase transition. 

%Through further investigation on the topology of the two phases separated by a critical point, whether the critical point is a topological phase transition can be confirmed. 

%In several models examined, the fixed point of the RG flow corresponds to the minimum of entanglement entropy in the parameter space. 

%The advantage of small $\chi$ is that different phases can already be captured, and it requires much less numerical effort. On the other hand, we find that the drawback of small $\chi$ implementation is that the obtained critical point will not be acurate, and consequently it is ambiguous to investigate any kind of critical exponents. 

\bibliography{Literature}

\end{document}